\documentclass[12pt]{article}

\usepackage{multicol}

\def\Red{}
\def\Black{}
\def\Blue{}

\newcommand{\text}{\rm}
\begin{document}

\title{\Red\textbf{Remarks on the BRST Cohomology of Supersymmetric Gauge Theories \Black}%
\vspace{4mm}\bigskip }
\author{V.E.R. Lemes$^{\mathrm{a}}$, M. Picariello$^{\mathrm{b}}$, M.S. Sarandy$^{%
\mathrm{a}}$ and S.P. Sorella$^{\mathrm{a}}$ \and \textbf{\ }\vspace{3mm} \\
$^{\mathrm{a}}\,${\small {\textit{UERJ - Universidade do Estado do Rio de
Janeiro,}}} \\
{\small {\textit{Rua S\~{a}o Francisco Xavier 524, 20550-013, }}}\\
{\small {\textit{Maracan\~{a}, Rio de Janeiro, Brazil.}}}\vspace{3mm}\\
$^{\mathrm{b}}\,${\small {\textit{Universit{\`{a}} degli Studi di Milano,
via  {\small Celoria 16, 20143 Milano }}}}\\
{\small {\textit{and }}}\\
{\small {\textit{INFN-Milano, Italy}}}}
\maketitle

\begin{abstract}\Blue
The supersymmetric version of the descent equations following from the
Wess-Zumino consistency condition is discussed. A systematic framework in
order to solve them is proposed. \ 
\Black
\newpage
\end{abstract}

\section{Introduction}

One of the most attractive properties of supersymmetric quantum field
theories is their softer ultraviolet behavior. Supersymmetry has allowed to
establish several nonrenormalization theorems \cite{nr}, which have provided
examples of gauge theories with vanishing beta function to all orders of
perturbation theory.

Recently, a criterion of general applicability for the ultraviolet
finiteness has been proven \cite{th}. The result allows to give a purely
cohomological algebraic characterization of the ultraviolet behavior of
gauge field theories, including the supersymmetric models as well. Moreover,
it also covers the case of theories whose beta function receives only
one-loop contribution, as it happens for the $N=2$ supersymmetric gauge
theories in four dimensions.

The aforementioned criterion makes use of the set of descent equations
stemming from the Wess-Zumino consistency condition. In the case of
supersymmetric theories it turns out that these equations take a peculiar
form, leading to a system of nonstandard equations which highly constrains
the possible invariant counterterms and anomalies allowed by the gauge
invariance and by the global supersymmetry. These equations have been proven
to be very useful in the algebraic proof of the ultraviolet finiteness
properties of both $N=2$ \cite{n2} and $N=4$ supersymmetric gauge theories 
\cite{n4,n41}.

It is worth mentioning that, unlike the nonsupersymmetric case, where
systematic procedures are available in order to solve the descent equations 
\cite{book,pr}, in the supersymmetric case the task is considerably more
difficult, even when a superspace formulation is available \cite{sp}.

The aim of this letter is to pursue the investigation of the structure of
the descent equations for supersymmetric gauge theories, by providing a
systematic framework to solve them.

The paper is organized as follows. In Sect.2 a short review of the
quantization of the supersymmetric gauge theories is given. In Sect.3 the
supersymmetric descent equations are discussed and a way to solve them is
presented. In Sect.4 the example of the $N=1$ supersymmetric Yang-Mills in
four dimensions is worked out.

\section{Algebraic structure of supersymmetric gauge theories}

A brief account of the quantization of the supersymmetric gauge theories in
the Wess-Zumino gauge is given here, following the procedure outlined in
refs.\cite{n4,mg,mp}. Let us start by considering a supersymmetric gauge
theory described by the classical action $\Sigma _{\mathrm{inv}}(\Phi )$,
where $\Phi $ denotes collectively the gauge and matter fields. In the
following we shall refer to renormalizable gauge theories in four
dimensions, the generalization to other dimensions being straightforward.

In the absence of central charges and adopting the Wess-Zumino gauge, the
supersymmetry algebra has the typical form 
\begin{eqnarray}
\left\{ Q_{\alpha }^{i},\overline{Q}_{\,\stackrel{.}{\alpha }}^{j}\right\}
&=&-2i\delta ^{ij}\,\sigma _{\alpha \stackrel{.}{\alpha }}^{\mu }\partial
_{\mu }+\mathrm{(gauge\,\,transf.)}+\mathrm{(eqs.\,of\,\,\,motion)}, 
\nonumber \\
\left\{ Q_{\alpha }^{i},Q_{\beta }^{j}\right\} &=&\left\{ \overline{Q}_{%
\stackrel{.}{\alpha }}^{i},\overline{Q}_{\stackrel{.}{\beta }}^{j}\right\} =%
\mathrm{(gauge\,\,transf.)}+\mathrm{(eqs.\,of\,\,\,motion)}  \label{salg}
\end{eqnarray}
where $Q_{\alpha }^{i}\,,\overline{Q}_{\,\stackrel{.}{\alpha }}^{j}$ are the
supersymmetric charges, with $\alpha ,\stackrel{.}{\alpha }\,=1,2$ being the
spinor indices and $i\,,\,j=1,...,N$ labelling the number of supersymmetries.

The action $\Sigma _{\mathrm{inv}}(\Phi )$ is left invariant by the charges $%
Q_{\alpha }^{i}\,,\overline{Q}_{\,\stackrel{.}{\alpha }}^{j}$. It is also
required to be invariant under gauge transformations, which give rise to the
nilpotent BRST\ operator $s$ when the local gauge parameter is replaced by
the Faddeev-Popov ghost. In order to properly quantize the theory, one has
to introduce the gauge-fixing and the antifield terms in the action. The
standard procedure in order to take into account both BRST and supersymmetry
invariance, is to collect them into a unique generalized BRST\ operator $Q$ 
\cite{n4,mg,mp}$.$ In addition to the Faddeev-Popov ghost, the introduction
of constant ghosts $\varepsilon _{i}^{\alpha },\overline{\varepsilon }_{j}^{%
\stackrel{.}{\alpha }}$ corresponding to global supersymmetry is required.
The resulting generalized operator $Q$ is found to be 
\begin{equation}
Q=s+\varepsilon _{i}^{\alpha }Q_{\alpha }^{i}+\overline{\varepsilon }_{j}^{%
\stackrel{.}{\alpha }}\ \overline{Q}_{\,\stackrel{.}{\alpha }}^{j}\;\;.
\label{q-op}
\end{equation}
The action $\Sigma _{\mathrm{inv}}(\Phi )$ is invariant under the $Q$%
-transformations which, due to the algebra $\left( \ref{salg}\right) $, turn
out to be nilpotent only up equations of motion and space-time translations,
namely 
\begin{equation}
Q^{2}=\varepsilon ^{\mu }\partial _{\mu }+\mathrm{(eqs.\,\,of\,\,motion)\;,}
\label{snil}
\end{equation}
with $\varepsilon ^{\mu }=$ $-2i\varepsilon _{i}^{\alpha }\sigma _{\alpha 
\stackrel{.}{\alpha }}^{\mu }\overline{\varepsilon }_{i}^{\stackrel{.}{%
\alpha }}.$

Hence, the complete classical action $\Sigma $ is given by 
\begin{equation}
\Sigma =\Sigma _{\mathrm{inv}}(\Phi )+\Sigma _{\mathrm{gf}}(\Phi ,\pi ,c,%
\overline{c})+\Sigma _{\mathrm{ext}}(\Phi ,\Phi ^{*},c,c^{*})\;,
\label{totac}
\end{equation}
where $\Sigma _{\mathrm{gf}}(\Phi ,\pi ,c,\overline{c})$ is the gauge-fixing
action depending on the gauge and matter fields $\Phi $, the Lagrange
multiplier $\pi ,$the Faddeev-Popov ghost $c$ and antighost $\overline{c}$.
The term $\Sigma _{\mathrm{ext}}(\Phi ,\Phi ^{*},c,c^{*})$ denotes the
antifields action, which is constructed by coupling the nonlinear $Q$%
-transformations to external fields $\Phi ^{*}$ and $c^{*}$ associated
respectively to $\Phi $ and $c$, \textit{i.e.} 
\begin{equation}
\Sigma _{\mathrm{ext}}(\Phi ,\Phi ^{*},c,c^{*})=\int d^{4}x\,\left(
\sum_{\Phi }\Phi ^{*}\,Q\Phi +c^{*}Qc+\left( \mathrm{terms\,\;quadratic\,%
\;in\,\;\,}\Phi ^{*}\,\mathrm{,}c^{*}\right) \right) \mathrm{.}
\label{extac}
\end{equation}
As is well known, the terms quadratic in the external fields $\left( \Phi
^{*},c^{*}\right) $ are needed in order to account for the on-shell
nilpotency of the generalized operator $Q$ \cite{n4,mg,mp}. The invariance
of the action $\Sigma _{\mathrm{inv}}(\Phi )\,$under $Q$ may be now
translated into the classical Slavnov-Taylor identity, whose typical form is 
\cite{vit,n2,n41} 
\begin{equation}
\mathcal{S}(\Sigma )=\varepsilon ^{\mu }\Delta _{\mu }^{\mathrm{cl}}\;,
\label{st}
\end{equation}
with 
\begin{equation}
\mathcal{S}(\Sigma )=\int d^{4}x\,\left( \sum_{\Phi }\frac{\delta \Sigma }{%
\delta \Phi ^{*}}\frac{\delta \Sigma }{\delta \Phi }+\frac{\delta \Sigma }{%
\delta c^{*}}\frac{\delta \Sigma }{\delta c}+Q\overline{c}\frac{\delta
\Sigma }{\delta \overline{c}}+Q\pi \frac{\delta \Sigma }{\delta \pi }\right)
\;.
\end{equation}
It is worth underlining that the breaking term $\Delta _{\mu }^{\mathrm{cl}}$
is a classical breaking, as it turns out to be linear in the quantum fields.
As such, it will not be affected at the quantum level \cite{book}.

Introducing the so called linearized Slavnov-Taylor operator $\mathcal{B}%
_{\Sigma }$ 
\begin{equation}
\mathcal{B}_{\Sigma }=\int d^{4}x\,\left( \sum_{\Phi }\left( \frac{\delta
\Sigma }{\delta \Phi ^{*}}\frac{\delta }{\delta \Phi }+\frac{\delta \Sigma }{%
\delta \Phi }\frac{\delta }{\delta \Phi ^{*}}\right) +\frac{\delta \Sigma }{%
\delta c^{*}}\frac{\delta }{\delta c}+\frac{\delta \Sigma }{\delta c}\frac{%
\delta }{\delta c^{*}}+Q\overline{c}\frac{\delta }{\delta \overline{c}}+Q\pi 
\frac{\delta }{\delta \pi }\right) \;,  \label{sto}
\end{equation}
it follows that 
\begin{equation}
\mathcal{B}_{\Sigma }\mathcal{B}_{\Sigma }=\varepsilon ^{\mu }\partial _{\mu
}\,\,,  \label{stonil}
\end{equation}
meaning that $\mathcal{B}_{\Sigma }$ is nilpotent only modulo a total
derivative. Of course, this property follows from the supersymmetric
structure of the theory. Moreover, the operator $\mathcal{B}_{\Sigma }$ is
strictly nilpotent when acting on the space of the integrated local
functionals of the fields, antifields and their derivatives. This is
precisely the functional space to which all invariant counterterms and
anomalies belong.

\section{The supersymmetric descent equations}

In order to discuss the structure of the supersymmetric descent equations,
let us begin by considering the Wess-Zumino consistency condition for the
invariant counterterms which can be freely added to any order of the
perturbation theory, namely 
\begin{equation}
\mathcal{B}_{\Sigma }\int d^{4}x\,\Omega ^{0}=0\;,  \label{w-z}
\end{equation}
where $\Omega ^{0}$ has the same quantum numbers of the classical
Lagrangian, \textit{i.e.} it is a local polynomial of dimension four and
vanishing Faddeev-Popov charge. The integrated consistency condition $\left( 
\ref{w-z}\right) $ can be translated at the local level as 
\begin{equation}
\mathcal{B}_{\Sigma }\Omega ^{0}=\partial ^{\mu }\Omega _{\mu \;}^{1},
\label{1-l}
\end{equation}
where $\Omega _{\mu \;}^{1}$is a local polynomial of Faddeev-Popov charge 1
and dimension 3. Applying now the operator $\mathcal{B}_{\Sigma }$ to both
sides of $\left( \ref{1-l}\right) $ and making use of eq.$\left( \ref{stonil}%
\right) $, one obtains the condition 
\begin{equation}
\partial ^{\mu }\left( \mathcal{B}_{\Sigma }\Omega _{\mu \;}^{1}-\varepsilon
_{\mu }\Omega ^{0}\right) =0\;,  \label{l-2}
\end{equation}
which, due to the algebraic Poincar\'{e} Lemma \cite{book}, implies 
\begin{equation}
\mathcal{B}_{\Sigma }\Omega _{\mu \;}^{1}=\varepsilon _{\mu }\Omega
^{0}+\partial ^{\nu }\Omega _{[\nu \mu ]}^{2}\;,  \label{l-3}
\end{equation}
for some local polynomial $\Omega _{[\nu \mu ]}^{2}$ antisymmetric in the
Lorentz indices $\mu ,\nu $ and with Faddeev-Popov charge 2. This procedure
can be easily iterated, yielding the following set of descent equations 
\begin{eqnarray}
\mathcal{B}_{\Sigma }\Omega ^{0} &=&\partial ^{\mu }\Omega _{\mu }^{1}\;, 
\nonumber \\
\,\,\,\,\,\,\,\,\;\;\;\;\;\mathcal{B}_{\Sigma }\Omega _{\mu }^{1}
&=&\partial ^{\nu }\Omega _{[\nu \mu ]}^{2}+\varepsilon _{\mu }\Omega ^{0}\;,
\nonumber \\
\mathcal{B}_{\Sigma }\Omega _{[\mu \nu ]}^{2} &=&\partial ^{\rho }\Omega
_{[\rho \mu \nu ]}^{3}+\varepsilon _{\mu }\Omega _{\nu }^{1}-\varepsilon
_{\nu }\Omega _{\mu }^{1}\;,  \nonumber \\
\mathcal{B}_{\Sigma }\Omega _{[\mu \nu \rho ]}^{3} &=&\partial ^{\sigma
}\Omega _{[\sigma \mu \nu \rho ]}^{4}+\varepsilon _{\mu }\Omega _{[\nu \rho
]}^{2}+\varepsilon _{\rho }\Omega _{[\mu \nu ]}^{2}+\varepsilon _{\nu
}\Omega _{[\rho \mu ]}^{2}\;,  \nonumber \\
\mathcal{B}_{\Sigma }\Omega _{[\mu \nu \rho \sigma ]}^{4} &=&\varepsilon
_{\mu }\Omega _{[\nu \rho \sigma ]}^{3}-\varepsilon _{\sigma }\Omega _{[\mu
\nu \rho ]}^{3}+\varepsilon _{\rho }\Omega _{[\sigma \mu \nu
]}^{3}-\varepsilon _{\nu }\Omega _{[\rho \sigma \mu ]}^{3}\;\,.  \label{sde}
\end{eqnarray}
It should be observed that these equations are of an unusual type, as the
cocycles with lower Faddeev-Popov charge appear in the equations of those
with higher Faddeev-Popov charge, turning the system $\left( \ref{sde}%
\right) $ highly nontrivial. We also remark that the last equation for $%
\Omega _{[\mu \nu \rho \sigma ]}^{4}$ is not homogeneous, a property which
strongly constrains the possible solutions. Eqs.$\left( \ref{sde}\right) $
immediately generalize to possible anomalies and to cocycles with arbitrary
Faddeev-Popov charge. To some extent, the system $\left( \ref{sde}\right) $
displays a certain similarity with the descent equations in $N=1$ superspace 
\cite{sp}. Actually, it is possible to solve the eqs.$\left( \ref{sde}%
\right) $ in a rather direct way by making use of the supersymmetric
structure of the theory. This goal is achieved by introducing an operator $%
\mathcal{W}_{\mu }$ which, together with the linearized Slavnov-Taylor
operator $\mathcal{B}_{\Sigma }$, gives rise to the algebra 
\[
\left\{ \mathcal{W}_{\mu },\mathcal{B}_{\Sigma }\right\} =\partial _{\mu
}\,\,\,,\,\,\,\left\{ \mathcal{W}_{\mu },\mathcal{W}_{\nu }\right\} =0\,\;%
\mathrm{.}
\]
The operator $\mathcal{W}_{\mu }$ has been introduced first in the case of
topological field theories \cite{tp,tp1}, and subsequently in the case of
extended supersymmetry \cite{th,n41}. In the next section the explicit form
of $\mathcal{W}_{\mu }$ for the case of $N=1$ gauge theories will be given.

Once the operator $\mathcal{W}_{\mu }$ has been introduced, it can be used
as a climbing operator for the descent equations $\left( \ref{sde}\right) .$
It turns out in fact that, provided an explicit form for $\Omega _{[\mu \nu
\rho \sigma ]}^{4}$ is available, a solution of the system is obtained by
repeated applications of $\mathcal{W}_{\mu }$ on $\Omega _{[\mu \nu \rho
\sigma ]}^{4}$, according to 
\begin{eqnarray}
\Omega ^{0} &=&\frac{1}{4!}\mathcal{W}^{\mu }\mathcal{W}^{\nu }\mathcal{W}%
^{\rho }\mathcal{W}^{\sigma }\Omega _{[\sigma \rho \nu \mu ]}^{4}\;, 
\nonumber \\
\Omega _{\mu }^{1} &=&\frac{1}{3!}\mathcal{W}^{\nu }\mathcal{W}^{\rho }%
\mathcal{W}^{\sigma }\Omega _{[\sigma \rho \nu \mu ]}^{4}\;,  \nonumber \\
\Omega _{[\mu \nu ]}^{2} &=&\frac{1}{2!}\mathcal{W}^{\rho }\mathcal{W}%
^{\sigma }\Omega _{[\sigma \rho \mu \nu ]}^{4}\;,  \nonumber \\
\Omega _{[\mu \nu \rho ]}^{3} &=&\mathcal{W}^{\sigma }\Omega _{[\sigma \mu
\nu \rho ]}^{4}\;.  \label{sol-desc}
\end{eqnarray}
We are left thus with the characterization of $\Omega _{[\mu \nu \rho \sigma
]}^{4}.$ This point can be faced by introducing a new operator $\mathcal{F}%
_{\Sigma }\;$defined as 
\begin{equation}
\mathcal{F}_{\Sigma }=\mathcal{B}_{\Sigma }-\varepsilon ^{\mu }\mathcal{W}%
_{\mu }\,\;.  \label{newb}
\end{equation}
Unlike $\mathcal{B}_{\Sigma },$ the new operator $\mathcal{F}_{\Sigma }$ has
the remarkable property of being strictly nilpotent, \textit{i.e.} 
\[
\mathcal{F}_{\Sigma }\mathcal{F}_{\Sigma }=0\,\,\,\,,\,\,\,\left\{ \mathcal{W%
}_{\mu },\mathcal{F}_{\Sigma }\right\} =\partial _{\mu }\;\,. 
\]
In particular, thanks to $\left( \ref{sol-desc}\right) $, the last equation
for $\Omega _{[\mu \nu \rho \sigma ]}^{4}$ in $\left( \ref{sde}\right) $ can
be cast in the form of a homogeneous equation 
\begin{equation}
\mathcal{F}_{\Sigma }\Omega _{[\mu \nu \rho \sigma ]}^{4}=0\;.  \label{h-e}
\end{equation}
This means that $\Omega _{[\mu \nu \rho \sigma ]}^{4}$ can be obtained from
the knowledge of the cohomology of the nilpotent operator $\mathcal{F}%
_{\Sigma }$, for which standard techniques are available \cite{pr}. This
gives us a systematic framework for solving the descent equations in the
supersymmetric case.

\section{The example of $N=1$ super Yang-Mills theory}

The $N=1$ super Yang-Mills action $S^{\mathrm{N=1}}$ in the Wess-Zumino
gauge is given by 
\begin{equation}
S^{\mathrm{N=1}}=\frac{1}{g^{2}}\mathrm{Tr}\int d^{4}x\left( -\frac{1}{4}%
F_{\mu \nu }F^{\mu \nu }-i\,\lambda ^{\alpha }\sigma _{\alpha \stackrel{.}{%
\beta }}^{\mu }D_{\mu }\overline{\lambda }^{\stackrel{.}{\beta }}+\frac{1}{2}%
\mathcal{D}^{2}\right) \;,  \label{n1ac}
\end{equation}
where $F_{\mu \nu }=\partial _{\mu }A_{\nu }-\partial _{\nu }A_{\mu }+\left[
A_{\mu },A_{\nu }\right] \,$is the field strength, $\lambda ^{\alpha }$ and $%
\overline{\lambda }^{\stackrel{.}{\beta }}$ are two-component spinors and $%
\mathcal{D}$ is an auxiliary scalar field, introduced for the off-shellness
closure of the supersymmetric algebra.

The action $S^{\mathrm{N=1}}$ is invariant under both BRST\ and
supersymmetry transformations. Following the general procedure, we shall
collect the BRST\ differential $s$ and the supersymmetry generators $\left(
Q_{\alpha },\overline{Q}_{\stackrel{.}{\alpha }}\right) $ into an extended
operator $Q$ 
\begin{equation}
Q=s+\varepsilon ^{\alpha }Q_{\alpha }+\overline{\varepsilon }^{\stackrel{.}{%
\alpha }}\overline{Q}_{\stackrel{.}{\alpha }}\ ,  \label{qn1}
\end{equation}
where $\varepsilon ^{\alpha }$and $\overline{\varepsilon }^{\stackrel{.}{%
\alpha }}$ are global ghosts. The operator $Q$ acts on the fields as 
\begin{eqnarray}
&&QA_{\mu }=-D_{\mu }c+\varepsilon ^{\alpha }\sigma _{\mu \,\alpha \stackrel{%
.}{\beta }}\,\overline{\lambda }^{\stackrel{.}{\beta }}+\lambda ^{\alpha
}\sigma _{\mu \,\alpha \stackrel{.}{\beta }}\,\overline{\varepsilon }^{%
\stackrel{.}{\beta }},  \nonumber \\
&&Q\lambda ^{\beta }=\left\{ c,\lambda ^{\beta }\right\} -\frac{1}{2}%
\varepsilon ^{\alpha }\left( \sigma ^{\mu \nu }\right) _{\alpha
}^{\,\,\,\,\beta }F_{\mu \nu }-\varepsilon ^{\beta }\mathcal{D},  \nonumber
\\
&&Q\overline{\lambda }^{\stackrel{.}{\beta }}=\left\{ c,\overline{\lambda }^{%
\stackrel{.}{\beta }}\right\} +\frac{1}{2}\left( \overline{\sigma }^{\mu \nu
}\right) _{\,\,\,\,\stackrel{.}{\alpha }}^{\stackrel{.}{\beta }}\,\overline{%
\varepsilon }^{\stackrel{.}{\alpha }}F_{\mu \nu }+\overline{\varepsilon }^{%
\stackrel{.}{\beta }}\mathcal{D},  \nonumber \\
&&Q\mathcal{D}=\left[ c,\mathcal{D}\right] -i\,\varepsilon ^{\alpha }\sigma
_{\alpha \stackrel{.}{\beta }}^{\mu }D_{\mu }\overline{\lambda }^{\stackrel{.%
}{\beta }}+i\,D_{\mu }\lambda ^{\alpha }\sigma _{\alpha \stackrel{.}{\beta }%
}^{\mu }\overline{\varepsilon }^{\stackrel{.}{\beta }},  \nonumber \\
&&Qc=c^{2}+2i\varepsilon ^{\alpha }\sigma _{\mu \,\alpha \stackrel{.}{\beta }%
}\,\overline{\varepsilon }^{\stackrel{.}{\beta }}A^{\mu }.  \label{qtransn1}
\end{eqnarray}
For the complete gauge-fixed action $\Sigma $ we have 
\begin{equation}
\Sigma =S^{\mathrm{N=1}}+S_{\mathrm{gf}}+S_{\mathrm{ext}}\;,  \label{totacn1}
\end{equation}
where $S_{\mathrm{gf}}$ is the gauge-fixing term in the Landau gauge and $S_{%
\mathrm{ext}}$ contains the coupling of the non-linear transformations $%
Q\Phi _{i}$ to the antifields $\Phi _{i}^{*}=(A_{\mu }^{*}$, $c^{*}$, $%
\lambda ^{\alpha \,*}$, $\overline{\lambda }_{\stackrel{.}{\alpha }}^{*}$, $%
\mathcal{D}^{*})$. They are given by 
\begin{eqnarray}
&&S_{\mathrm{gf}}=\mathrm{Tr}\int d^{4}x\,Q\left( \bar{c}\partial A\right) \;%
\mathrm{,}  \nonumber \\
&&S_{\mathrm{ext}} =\mathrm{Tr}\int d^{4}x\left( A_{\mu }^{*}QA^{\mu
}+c^{*}Qc+\lambda ^{\alpha \,*}Q\lambda _{\alpha }+\overline{\lambda }_{%
\stackrel{.}{\alpha }}^{*}Q\overline{\lambda }^{\stackrel{.}{\alpha }}+%
\mathcal{D}^{*}Q\mathcal{D}\right),  \label{reston1}
\end{eqnarray}
with $Q\bar{c}=b\,\,$ and $\,\,Qb=-2i\varepsilon ^{\alpha }\sigma _{\alpha 
\stackrel{.}{\beta }}^{\mu }\,\overline{\varepsilon }^{\stackrel{.}{\beta }%
}\partial _{\mu }\bar{c}$.

As usual, $\bar{c},b$ denote the antighost and the Lagrange multiplier. The
operator $Q$ turns out to be nilpotent only up to space-time translations 
\begin{equation}
Q^{2}=-2i\varepsilon ^{\alpha }\sigma _{\alpha \stackrel{.}{\beta }}^{\mu }\,%
\overline{\varepsilon }^{\stackrel{.}{\beta }}\partial _{\mu }\;.
\label{q2n1}
\end{equation}
The complete action $\Sigma $ satisfies the following Slavnov-Taylor
identity 
\begin{equation}
\mathcal{S}(\Sigma )=-2i\varepsilon ^{\alpha }\sigma _{\alpha \stackrel{.}{%
\beta }}^{\mu }\,\overline{\varepsilon }^{\stackrel{.}{\beta }}\Delta _{\mu
}^{\mathrm{cl}}\;,  \label{n1slav}
\end{equation}
where 
\begin{equation}
\mathcal{S}(\Sigma )=\mathrm{Tr}\int d^{4}x\left( \frac{\delta \Sigma }{%
\delta \Phi _{i}^{*}}\frac{\delta \Sigma }{\delta \Phi _{i}}+b\frac{\delta
\Sigma }{\delta \bar{c}}-2i\varepsilon ^{\alpha }\sigma _{\alpha \stackrel{.%
}{\beta }}^{\mu }\,\overline{\varepsilon }^{\stackrel{.}{\beta }}\partial
_{\mu }\bar{c}\frac{\delta \Sigma }{\delta b}\right) \;  \label{n1slav2}
\end{equation}
and the classical breaking $\Delta _{\mu }^{\mathrm{cl}}$ is 
\begin{equation}
\Delta _{\mu }^{\mathrm{cl}}=\mathrm{Tr}\int d^{4}x\left( -A^{*\nu }\partial
_{\mu }A_{\nu }+\frac {} {}c^{*}\partial _{\mu }c+\lambda ^{\alpha
\,*}\partial _{\mu }\lambda _{\alpha }+\overline{\lambda }_{\stackrel{.}{%
\alpha }}^{*}\partial _{\mu }\overline{\lambda }^{\stackrel{.}{\alpha }}-%
\mathcal{D}^{*}\partial _{\mu }\mathcal{D}\right) .  \label{qcn1}
\end{equation}
From eq.$\left( \ref{n1slav}\right) $ and $\left( \ref{n1slav2}\right) $ it
follows that the linearized operator $B_{\Sigma }$ defined as 
\begin{equation}
\mathcal{B}_{\Sigma }=\mathrm{Tr}\int d^{4}x\left( \frac{\delta \Sigma }{%
\delta \Phi _{i}^{*}}\frac{\delta }{\delta \Phi _{i}}+\frac{\delta \Sigma }{%
\delta \Phi _{i}}\frac{\delta }{\delta \Phi _{i}^{*}}+b\frac{\delta }{\delta 
\bar{c}}-2i\varepsilon ^{\alpha }\sigma _{\alpha \stackrel{.}{\beta }}^{\mu
}\,\overline{\varepsilon }^{\stackrel{.}{\beta }}\partial _{\mu }\bar{c}%
\frac{\delta }{\delta b}\right) 
\end{equation}
is nilpotent modulo a total space-time derivative, namely 
\begin{equation}
\mathcal{B}_{\Sigma }\mathcal{B}_{\Sigma }=\varepsilon ^{\alpha }d_{\alpha
}\;\,,  \label{niln1}
\end{equation}
with the operator $d_{\alpha }$ given by 
\begin{equation}
d_{\alpha }=-2i\sigma _{\alpha \stackrel{.}{\beta }}^{\mu }\,\overline{%
\varepsilon }^{\stackrel{.}{\beta }}\partial _{\mu }\,.  \label{dan1}
\end{equation}
The integrated cohomology of $\mathcal{B}_{\Sigma }$ is characterized by a
consistency condition of the kind $\left( \ref{w-z}\right) $ which, in the
present case, can be written as \cite{ul} 
\begin{eqnarray}
\mathcal{B}_{\Sigma }\Omega ^{0} &=&d_{\alpha }\Omega ^{1\alpha }\;, 
\nonumber \\
\,\,\,\,\,\,\,\,\;\;\;\;\;\mathcal{B}_{\Sigma }\Omega ^{1\alpha }
&=&d_{\beta }\Omega ^{2[\beta \alpha ]}+\varepsilon ^{\alpha }\Omega ^{0}\;,
\nonumber \\
\mathcal{B}_{\Sigma }\Omega ^{2[\beta \alpha ]} &=&\varepsilon ^{\beta
}\Omega ^{1\alpha }-\varepsilon ^{\alpha }\Omega ^{1\beta }\;.  \label{den1}
\end{eqnarray}
It should be noted that the presence of the operator $d_{\alpha }$ in the
first equation of $\left( \ref{den1}\right) $ is due to the supersymmetric
character of the theory, following by observing that in the
nonsupersymmetric case the pure Yang-Mills Lagrangian \textrm{Tr}$\left(
F_{\mu \nu }F^{\mu \nu }\right) $ is pointwise invariant.

Defining now the climbing operator $\mathcal{W}_{\alpha }$ 
\begin{equation}
\mathcal{W}_{\alpha }=\left[ \frac{\partial }{\partial \varepsilon ^{\alpha }%
},B_{\Sigma }\right] \;,  \label{con1}
\end{equation}
it is easily verified that 
\begin{eqnarray}
\left\{ \mathcal{W}_{\alpha },B_{\Sigma }\right\} = d_{\alpha }\,\,\, ,
\,\,\, \left\{ \mathcal{W}_{\alpha },\mathcal{W}_{\beta }\right\} =0\,\;%
\mathrm{.}  \label{algn1}
\end{eqnarray}
As discussed in the previous section, the next step is the introduction of
the operator $\mathcal{F}_{\Sigma }\;$%
\begin{equation}
\mathcal{F}_{\Sigma }=\mathcal{B}_{\Sigma }-\varepsilon ^{\alpha }\mathcal{W}%
_{\alpha \;}.  \label{nbn1}
\end{equation}
Accordingly, the last equation of $\left( \ref{den1}\right) $ reads 
\begin{equation}
\mathcal{F}_{\Sigma }\Omega ^{2[\beta \alpha ]}=0\;.  \label{le}
\end{equation}
Furthermore, up to trivial exact cocycles, $\Omega ^{2[\beta \alpha ]}$ is
found to be 
\begin{equation}
\Omega ^{2[\beta \alpha ]}=\varepsilon ^{\beta \alpha }\,\,\mathrm{Tr\,\,}%
\lambda ^{\gamma }\lambda _{\gamma }\;.  \label{botn1}
\end{equation}
The higher cocycles are obtained by applying repeatedly the operator $%
\mathcal{W}_{\alpha }$ on $\Omega ^{2[\beta \alpha ]}$ 
\begin{eqnarray}
\Omega ^{0} &=&\frac{1}{2}\mathcal{W}_{\alpha }\mathcal{W}_{\beta }\Omega
^{2[\beta \alpha ]}\;,  \nonumber \\
\Omega ^{1\alpha } &=&\mathcal{W}_{\beta }\Omega ^{2[\beta \alpha ]}\;.
\label{soln1}
\end{eqnarray}
Acting now with $\partial /\partial g$ on both sides of the Slavnov-Taylor
identity $\left( \ref{n1slav}\right) $ and observing that the linear
breaking term $\Delta _{\mu }^{\mathrm{cl}}$ does not depend on the coupling
constant $g$, we get the condition 
\begin{equation}
\mathcal{B}_{\Sigma }\frac{\partial \Sigma }{\partial g}=0\;,  \label{n2cond}
\end{equation}
which shows that $\partial \Sigma /\partial g$ is invariant under the action
of $\mathcal{B}_{\Sigma }$. In fact $\partial \Sigma /\partial g$ identifies
the cohomology of $\mathcal{B}_{\Sigma }$ in sector of the integrated
polynomials with dimension four and ghost number zero, belonging to the same
cohomology class of $\,\int d^{4}x\,\Omega ^{0}.$

From eqs.$\left( \ref{soln1}\right) ,$ the usefulness of the operator $%
\mathcal{W}_{\mu }$ becomes now apparent. In particular, it allows to
establish the following relation 
\begin{equation}
\frac{\partial \Sigma }{\partial g}=\frac{1}{4g^{3}}\varepsilon ^{\alpha
\beta }\,\,\mathcal{W}_{\alpha }\mathcal{W}_{\beta }\,\mathrm{Tr\,\,}\int
d^{4}x\lambda ^{\gamma }\lambda _{\gamma }\;\,.  \label{acn1final}
\end{equation}
Equation $\left( \ref{acn1final}\right) $ implies that the origin of the
action of $N=1$ super Yang-Mills can be traced back to the gauge invariant
local polynomial $\mathrm{Tr\,\,}\int d^{4}x\lambda ^{\gamma }\lambda
_{\gamma }.\;$This relationship has recently been pointed out in \cite{ul}.

The construction of the nilpotent operator $\mathcal{F}_{\Sigma }$ of
equation $\left( \ref{newb}\right) $ is easily generalized to the cases of $%
N=2$ and $N=4$ gauge theories, so that the analogue of the equation $\left( 
\ref{acn1final}\right) $ can be worked out from the knowledge of its
cohomology, a representative of which has been given in \cite{n2,n41}.

\section{Conclusion}

The structure of the descent equations for supersymmetric gauge theories has
been discussed. Due to the supersymmetry algebra $\left( \ref{salg}\right) ,$
these equations are of an unusual type, a property which makes their
analysis rather cumbersome. However, it has been shown that a suitable
climbing operator $\mathcal{W}_{\mu }$ can be introduced by making use of
the proper supersymmetric algebra. Provided the solution $\Omega _{[\mu \nu
\rho \sigma ]}^{4}$ of the last equation of the system $\left( \ref{sde}%
\right) $ is available, a solution of the whole system is obtained by
repeated applications of $\mathcal{W}_{\mu }$ on $\Omega _{[\mu \nu \rho
\sigma ]}^{4}$. Concerning the characterization of $\Omega _{[\mu \nu \rho
\sigma ]}^{4}$, we have been able to prove that it belongs to the cohomology
of the nilpotent operator to $\mathcal{F}_{\Sigma }$ of eq.$\left( \ref{newb}%
\right) .$ As a consequence, it can be determined by standard cohomolgy
arguments \cite{book,pr}, providing thus a systematic framework for
analysing the supersymmetric version of the descent equations.

\section*{Acknowledgements}

The Conselho Nacional de Desenvolvimento Cient\'{\i }fico e Tecnol\'{o}gico
CNPq-Brazil, the Funda{\c{c}}{\~{a}}o de Amparo {\`{a}} Pesquisa do Estado
do Rio de Janeiro (Faperj), the SR2-UERJ and the MIUR (Ministero
dell'Istruzione, dell'Universit\`{a} e della Ricerca) are acknowledged for
the financial support. 

M. Picariello is grateful to the Theoretical Physics Department of UERJ for
kind hospitality.

\end{document}